\documentclass[11pt]{article}
\usepackage{amssymb,amsfonts}
\usepackage{amsmath,amsthm,amscd}
\usepackage[mathscr]{eucal}
\usepackage{color}
\usepackage{yfonts,mathrsfs}
\begin{document}

\renewenvironment{proof}{\begin{trivlist}\item[]
{\quad\, P r o o f. }}{\hfill\rule{0.5em}{0.5em}\end{trivlist}}
\newtheorem{theorem}{\,\quad Theorem}              
\newtheorem{lemma}{\,\quad Lemma}                  
\newtheorem{definition}{\,\quad Definition}        
\newtheorem{corollary}{\,\quad Corollary}          
\newtheorem{proposition}{\,\quad Proposition}      
\newtheorem{example}{\,\quad\rm E x a m p l e }    
\newtheorem{remark}{\,\quad\rm R e m a r k }       
\renewcommand{\thesubsection}{\arabic{subsection}.}
\renewcommand{\theequation}{\arabic{equation}}
\renewcommand{\thetheorem}{\arabic{theorem}}
\renewcommand{\thelemma}{\arabic{lemma}}
\renewcommand{\thedefinition}{\thesubsection\arabic{definition}}
\renewcommand{\thecorollary}{\thesubsection\arabic{corollary}}
\renewcommand{\theproposition}{\thesubsection\arabic{proposition}}
\renewcommand{\theremark}{\thesubsection\arabic{remark}}
\renewcommand{\theexample}{\thesubsection\arabic{example}}
\title{A formula for eigenvalues of Jacobi matrices \\with a reflection symmetry}
\author{ S.~B.~Rutkevich 
\vspace{.4cm}
\\
\small{Duisburg, Germany}}
\date{October 16,  2018} 
\maketitle

\section*{Abstract} The spectral properties of { two} special classes of Jacobi operators are studied. For the first 
class represented by the  $2M$-dimensional  real   Jacobi matrices { whose entries are symmetric with respect to
the secondary diagonal}, a 
 new polynomial identity relating the eigenvalues of such matrices with their matrix { entries} is obtained. In the limit
 $M\to\infty$ this identity induces some requirements, which should satisfy the scattering data of the
resulting infinite-dimensional Jacobi operator in the half-line, 
which super- and sub-diagonal matrix elements are equal to $-1$.  We obtain such requirements 
in the simplest case of the discrete Schr\"odinger operator acting in ${l}^2( \mathbb{N})$,
which does not have bound and semi-bound states, and which 
potential has a compact support.

\vskip2mm

\section{Introduction \label{intr}}
The theory of tridiagonal (Jacobi) matrices has rich applications in different fields of physics and mathematics including the theory of Anderson localization 
\cite{LGP,ChSin89,Aiz93}, { the} thermodynamic 
Casimir effect  \cite{DGHHRS14,RD15}, orthogonal polynomials \cite{deiftorthogonal}, and nonlinear integrable lattice models \cite{toda1989theory,teschl2000jacobi,Nov90}. The spectral properties of Jacobi matrices 
are to much extent similar to the spectral properties of the continuous one-dimensional Sturm-Liouville operators.  
For a detailed review  of the present stage of the theory of Jacobi operators see the monograph by G.~Teschl \cite{teschl2000jacobi}.

In this paper we address the spectral properties of a special class of Jacobi matrices, which are symmetric with respect
to both main and secondary diagonals. To be more specific, let us introduce the real Jacobi matrix, the Hamiltonian
\begin{equation}\label{HH}
H = \left(
\begin{matrix}
b_1	& a_1 		& 0		  		& 0		  		& 0 			\\
a_1	& b_2		& \ddots  		& 0		  		& 0 			\\
0 			& \ddots	  	& \ddots    	& \ddots 		& 0 			\\
0			& 0 	    	& a_{N - 2} 	& b_{N-1}	& a_{N-1} 	\\
0			& 0				& 0				& a_{N-1}	& b_N 	\\
\end{matrix}\right),
\end{equation}
which acts in the Hilbert space ${\mathbb C}^N$. Throughout this paper, the dimension of this space 
is supposed to be even,  
\begin{equation}\label{2M}
N=2 M. 
\end{equation}
The (real) entries of the matrix $H$ will be subject to the following 
constrains
\begin{eqnarray}\label{even}
b_{{N}+1-j}&=&b_j, \quad j=1,\ldots N,\\
a_{{N}-j}&=&a_j, \quad j =1,\ldots N-1.\nonumber
\end{eqnarray}
Jacobi matrices satisfying this symmetry property were studied by Hochstadt \cite{HOCHSTADT1974435}.

Since the Hamiltonian $H$ commutes with the reflection operator 
\begin{equation}\label{refl}
R: \Psi(j)\to \Psi(N+1-j),
\end{equation}
 where $\{\Psi(j)\}_{j=1}^N\in {\mathbb C}^N$, 
the eigenvectors $\{\Psi_n\}_{n=1}^N$ of the matrix \eqref{HH} can be classified by their parities
$\kappa_n=\pm1$,
\begin{eqnarray}\label{Schro}
H \,\Psi_n=\lambda_n\,\Psi_n,\\
 R \,\Psi_n=\kappa_n\,\Psi_n{.}
\end{eqnarray}
There are exactly $M$ eigenvectors $\Psi_m^{(ev)}$
of the matrix $H$, which are even  under the action of the 
reflection  \eqref{refl}, and $M$ eigenvectors  $\Psi_m^{(od)}$, which are odd. 
We shall use different notation $\mu_m$ and $\nu_n$ for the  corresponding 
eigenvalues, 
\begin{eqnarray}\label{eigv}
H\,\Psi_m^{(ev)}=\mu_m\,\Psi_m^{(ev)}, &\quad& \Psi_m^{(ev)}(N+1-j)=\Psi_m^{(ev)}(j),\\
H\,\Psi_m^{(od)}=\nu_m\,\Psi_m^{(od)},&\quad& \Psi_m^{(od)}(N+1-j)=-\Psi_m^{(od)}(j),\nonumber
\end{eqnarray}
with $j=1,\ldots,N$, $m=1,\ldots,M$.
 
As the main result, we prove the following 
\begin{theorem}\label{Th2}
Let $H$ be the real Jacobi matrix \eqref{HH} satisfying \eqref{2M}, and \eqref{even}.
 Then, the  eigenvalues $\{\mu_m\}_{m=1}^M$
and  $\{\nu_m\}_{m=1}^M$ of the matrix $H$ specified by relations \eqref{eigv} obey the equality
\begin{equation}\label{pr0}
\prod_{m=1}^M\prod_{n=1}^M(\mu_m-\nu_n)= (-1)^{M(M-1)/2}\,(2 a_M)^M \prod_{j=1}^{M-1}a_j^{2j}.
\end{equation}
\end{theorem}

Proceeding then to the limit $M\to\infty$, we obtain from \eqref{pr0}  identities \eqref{EQ}, \eqref{Jost} for the scattering data corresponding
to  the
discrete Schr\"odinger operator (i.e., the Jacobi operator acting in ${l}^2( \mathbb{N})$, which is 
described by the 
infinite-dimensional matrix \eqref{HH}  with $a_j=-1$ and $b_j=2+v_j$, $j\in \mathbb{N}$) under certain assumptions
on the potential $\{v_j\}_{j\in \mathbb{N}}$. These results are described in Section 4 and Theorem \ref{Th3} therein.

A part of the results described here were given in the  preprint 
\cite{R2014}.
\section{Proof of Theorem \ref{Th2} } \label{SLP}
First, let us prove the Theorem for the particular choice of the Jacobi matrix \eqref{HH},
which is  specified by the relations
 \begin{equation}\label{examp}
 a_j=-1,\quad  b_{j'}=2 \quad {\rm for\;all}\quad j=1,\ldots, N-1, \quad j'=1,,\ldots, N.
 \end{equation}
In this case, the solution of the eigenvalue problem \eqref{Schro} is well known,
\begin{eqnarray}\nonumber
\Psi_n(j)=\sin(k_n j),\\\nonumber
\lambda_n= \omega(k_n),\\
k_n=\frac{\pi n}{N+1}, \label{kn}\label{kn}
\end{eqnarray}
with  $n=1,\ldots,N$, and 
\begin{equation}\label{omega}
\omega(p)=4\sin^2(p/2). 
\end{equation}
This yields  
\begin{equation}\label{munu0}
\mu_m= 4 \sin^2\frac{ (2m-1)\pi}{2(2M+1)} , \quad
\nu_m=  4 \sin^2\frac{2 m \pi}{2(2M+1)},
\end{equation}
and equality \eqref{pr0}  reduces to the form
\begin{equation}\label{pr2}
\prod_{m=1}^M\prod_{n=1}^M\left[4 \sin^2\frac{ (2m-1)\pi}{2(2M+1)}-
 4 \sin^2\frac{2 n \pi}{2(2M+1)}\right]=2^M\,(-1)^{M(M+1)/2}.
\end{equation}
Proof of this equality is given in Appendix.

Turning to the eigenvalue problem for a general Jacobi matrix specified in the Theorem, we first 
rewrite it as a  difference equation
\begin{eqnarray} \label{eig}
&&a_j \,\Psi_n(j+1)+b_j \, \Psi_n(j) +a_{j-1} \, \Psi_n(j-1) =\lambda_n \,\Psi_n(j),  \\\label{sym}
&&j=1,\ldots,N,\quad n=1,\ldots,N,
\end{eqnarray}
subjected to constrains \eqref{even} and  supplemented with  the Dirichlet boundary conditions 
\begin{equation}\label{BC}
\Psi_n(0)= \Psi_n(N+1)=0.
\end{equation}

Let us  consider  two associated eigenvalue problems for the even and odd states, which
are restricted to the $M$-dimensional quotient vector space ${\mathbb C}^M$.  We shall use the
lower case 
$\psi$ to denote  vectors in this quotient space, $\{\psi(j)\}_{j=1}^M\in {\mathbb C}^M$. Of course, 
$\Psi(j)=\psi(j)$ for $j=1,\ldots, M$.

The  vectors $\psi_{m}^{(ev)}(j)$, $j=1,\ldots, M$,  are the eigenstates
of the tridiagonal $M\times M$ matrix $H^{(ev)}$:
\begin{eqnarray}\label{Hev}
H^{(ev)}=\left(\begin{array}{ccccccc}
b_1 &a_1&0&0&0&\dots&0\\
a_1&b_2&a_2&0&0&\dots&0\\
0&a_2&b_3&a_3&0&\dots&0\\
\dots&\dots&\dots&\dots&\dots&\dots&\dots\\
0&0&\dots&0&a_{M-2}&b_{M-1}&a_{M-1}\\
0&0&\dots&0&0&a_{M-1}&b_M+a_M
\end{array}\right),
\end{eqnarray}
{ corresponding to} eigenvalues $\mu_m$, $m=1,\ldots,M$.
Similarly, the
vectors $\psi_{m}^{(od)}(j)$, $j=1,\ldots, M$,  are the eigenstates
of the tridiagonal $M\times M$ matrix $H^{(od)}$:
\begin{eqnarray}\label{Hod}
H^{(od)}=\left(\begin{array}{ccccccc}
b_1 &a_1&0&0&0&\dots&0\\
a_1&b_2&a_2&0&0&\dots&0\\
0&a_2&b_3&a_3&0&\dots&0\\
\dots&\dots&\dots&\dots&\dots&\dots&\dots\\
0&0&\dots&0&a_{M-2}&b_{M-1}&a_{M-1}\\
0&0&\dots&0&0&a_{M-1}&b_M-a_M
\end{array}\right),
\end{eqnarray}
{ corresponding to} eigenvalues $\nu_m$, $m=1,\ldots,M$.
Note, that the matrices $H^{(ev)}$ and $H^{(od)}$ are simply related
\begin{equation} \label{P}
H^{(od)}-H^{(ev)}=-2\,a_M P,
\end{equation}
with the projecting  matrix  $P_{m,m'}=\delta_{m,M}\delta_{m',M}$,
$\;\;m,m'=1,\ldots,M$, and ${\rm rank}\,P=1$.

It is convenient now to allow parameters  $\{a_j,b_j\}_{j=1}^M$ in the matrices \eqref{Hev},  \eqref{Hod}  
to take complex values.
\begin{lemma}  \label{L1}The matrices $H^{(od)}$ and $H^{(ev)}$ defined by \eqref{Hev}, \eqref{Hod} \label{Le}
have no common eigenvalues for arbitrary complex
numbers $\{b_j\}_{j=1}^M$, if  $a_j\ne0$ for all $j=1,\ldots,M$.
\end{lemma}
\begin{proof}
{For a contradiction,} we shall assume  that the matrices $H^{(ev)}$ and $H^{(od)}$ have a common
eigenvalue $\Lambda$.

So, let us suppose that
\[
\sum_{j'=1}^M H^{(ev)}_{j,j'}x_{j'}=\Lambda\, x_j, \quad
\sum_{j'=1}^M H^{(od)}_{j,j'}y_{j'}=\Lambda \,y_j,
\]
with nonzero vectors $\{x_j\}_{j=1}^M$, $\{y_j\}_{j=1}^M$.
We get
\begin{eqnarray*}
 \Lambda \sum_{j=1}^M y_j\,x_j=\sum_{j=1}^M \sum_{j'=1}^M y_jH^{(ev)}_{j,j'}x_{j'}=
\sum_{j=1}^M \sum_{j'=1}^M y_j(H^{(od)}_{j,j'}+2 a_M\,P_{j,j'})x_{j'}=\\
\sum_{j=1}^M \sum_{j'=1}^M y_jH^{(od)}_{j,j'}x_{j'}
+2a_M\sum_{j=1}^M \sum_{j'=1}^M y_jP_{j,j'}x_{j'}=2 a_M\,y_M\,x_M+\Lambda \sum_{j=1}^M y_j\,x_j.
\end{eqnarray*}
Here we have taken into account, that the matrix $H^{(od)}$ is symmetric.
Therefore, 
\[
a_M\,y_M\,x_M=0.
\]
Since $a_M\ne0$, this means that at least one of the numbers $y_M$ and $x_M$ is zero.
However, if $y_M=0$, we conclude immediately\begin{footnote} 
{Really, if $0=y_M\equiv \psi(j=M)\equiv\Psi(j=M)$, then $\Psi(j=M+1)=-\Psi(j=M)=0$,
since $\Psi(j)=-\Psi(2M+1-j)$ for all $j=1,\dots,2M$.
And since the wave-function $\Psi(j)$ takes zero values at two neighbor
sites $\Psi(M)=\Psi(M+1)=0$, and $a_j\ne 0$ for all $j=1,\ldots,M$, 
one can check recursively from (\ref{eig}),
that $\Psi(M-1)=0, \;\Psi(M-2)=0, \;\dots,\; \Psi(1)=0$,
and, therefore, $\psi(j)=0$ for all $j=1,\dots,M$.}
\end{footnote}, that $y_m=0$ for all $m=1,\ldots,M$. 
Therefore,  $\Lambda$ {\it is not} an eigenvalue of $H^{(od)}$.
Similarly, if $x_M=0$, we conclude, that $x_m=0$ for all $m=1,\ldots,M$,
and, hence,  $\Lambda$ {\it is not} an eigenvalue of $H^{(ev)}$.
This contradiction  proves the Lemma.
\end{proof}

The  eigenvalues $\{\mu_m\}_{m=1}^M$ and $\{\nu_m\}_{m=1}^M$ are the zeroes of the characteristic polynomials
\begin{equation}
P^{(ev)}(\mu)=\det(\mu-H^{(ev)}),
\end{equation}
 and 
 \begin{equation}
P^{(od)}(\nu)=\det(\nu-H^{(od)}),
\end{equation}
 respectively. It follows from the Vieta's  theorem, that 
the symmetric polynomials of eigenvalues $\{\mu_m\}_{m=1}^M$,
as well as the symmetric polynomials of eigenvalues $\{\nu_m\}_{m=1}^M$,
can be written as polynomial functions of the variables $\{a_j\}_{j=1}^M$ and $\{b_j\}_{j=1}^M$.
This is true, in particular for the function
\begin{equation}\label{pr1}
\prod_{m=1}^M\prod_{n=1}^M(\mu_m-\nu_{n})=\tilde{Q}(a,b),
\end{equation}
since the product { on} the left-hand side is a symmetric polynomial of $\{\mu_m\}_{m=1}^M$, and it is 
also a symmetric polynomial of $\{\nu_m\}_{m=1}^M$. 

However, the polynomial $\tilde{Q}(a,b)$ in the 
right-hand side of \eqref{pr1} \underline{must not} depend 
on $\{b_j\}_{j=1}^M$, i.e. 
\begin{equation}\label{Qab}
\tilde{Q}(a,b)\equiv Q(a).
\end{equation}

 To prove this, let us fix all $a_j$   at 
some arbitrary nonzero complex values $a_j^{(0)}\in  {\mathbb C}$, $a_j^{(0)}\ne 0$ for all $j=1,\ldots,M$. Then
equation \eqref{pr1} takes the form 
\begin{eqnarray}\label{PR2}
&&\prod_{m=1}^M\prod_{n=1}^M[\mu_m(b)-\nu_{n}(b)]=P(b),\\
&&P(b)\equiv \tilde{Q}(a^{(0)},b),\nonumber
\end{eqnarray}
where we have explicitly indicated the algebraic dependence of the eigenvalues $\mu_m$ and $\nu_m$ on the parameters $b_j$.
One can easily see from \eqref{PR2} that the   polynomial $P(b)$ has no zeros. 
Really, if $P(b)=0$ at some  $b=b^{(0)}$, then at least one eigenvalue $\mu_{m_0}(b^{(0)})$ of the matrix  $H^{(ev)}(b^{(0)})$
must coincide with some eigenvalue $\nu_{n_0}(b^{(0)})$ of the matrix  $H^{(od)}(b^{(0)})$, as it is implied 
by equality \eqref{PR2}. However, this  is { a} contradiction with  Lemma \ref{L1}, which guaranties that matrices 
$H^{(ev)}(b^{(0)})$  {and} $H^{(od)}(b^{(0)})$ have no common eigenvalues. Since the polynomial $P(b)$ has no zeroes, it is just a constant, and we arrive { at} equation \eqref{Qab}.

Thus, we have proven that
\begin{equation}\label{Qa}
\prod_{m=1}^M\prod_{n=1}^M(\mu_m-\nu_{n})={Q}(a),
\end{equation}
with some polynomial $Q(a)$.
In order to find this  polynomial explicitly, let us address some { of}
its properties. 
\begin{lemma}
(a)
The polynomial $Q(a)$ defined by \eqref{Qa}  is homogeneous, and its 
 degree equals to $M^2$:
 \begin{equation}\label{Qun}
Q(t\, a)=t^{M^2}\,Q(a).
\end{equation}
(b) The polynomial $Q(a)$ can be represented in the form
\begin{equation}\label{Qa1}
Q(a)=a_1^2 a_2^4\ldots a_{M-1}^{2(M-1)}\,a_M^M\, Q_0(a),
\end{equation}
where $Q_0(a)$ is also some polynomial of $\{a_j\}_{j=1}^M$.
\end{lemma}
\begin{proof}
The statement (a) follows immediately from \eqref{Qa}. To prove (b), let us, first, note, that 
$Q(a)$ vanishes, if one of the variables $a_j$ takes the zero value. Really, if 
{$a_j=0$} for some 
$j$, 
the matrices $H^{(ev)}$ and $H^{(od)}$ defined by \eqref{Hev}, \eqref{Hod} become block-diagonal 
containing two blocks of dimensions $j$ and $M-j$. Furthermore, the $j$-dimensional blocks in their
top-left corners are identical. Therefore, $j$ eigenvalues of the matrix $H^{(ev)}$ merge 
with $j$ eigenvalues  of the matrix $H^{(od)}$ in this case, 
\begin{equation}\label{Lam}
\mu_m(a_j=0)=\nu_m(a_j=0)\equiv \Lambda_m , \;\; {\rm for} \;\; m=1,\ldots,j.
\end{equation}

First, let us {choose $j=1$, and} vary the parameter $a_1$ near the zero point $a_1\to0$ and keeping the rest $a_{j{'}}$, $j{'}=2,\dots,M $ at some fixed
non-zero values. Then, a simple perturbative analysis of equations 
\begin{eqnarray}\label{det}
\det(\mu-H^{(ev)})=0,\quad 
\det(\nu-H^{(od)})=0,
\end{eqnarray}
 yields  for small $a_1$:
\begin{eqnarray}\label{mum1}
\mu_1(a_{1})=
\Lambda_1+O(a_1^2), \\
\nu_1(a_1)=
\Lambda_1+O(a_1^2).\nonumber
\end{eqnarray}
Combining this result with \eqref{Qa}, we conclude that the polynomial $Q(a)$ has 
at least the second order zero {at} $a_1$.

Second, let us {choose $j=2$, and}  tune the  parameter $a_2$ near the origin $a_2\to0$ keeping all others $a_j$ nonzero.  
Then, for $j=1,2$, one can easily obtain from \eqref{Lam}, \eqref{det}
\begin{eqnarray}\label{mum2}
&&\mu_j(a_2)=
\Lambda_j +O(a_2^2), \\
&&\nu_j(a_2)=\nonumber
\Lambda_j+O(a_2^2).
\end{eqnarray}
Therefore,  two brackets $(\mu_1-\nu_1)$ and $(\mu_2-\nu_2)$  in the product { on} the left-hand side of 
\eqref{Qa} vanish at $a_2\to0$. Since the both brackets are of the order $O(a_2^2)$, the polynomial 
$Q(a)$ must have at least the fourth order zero {at} $a_2$.

Finally, if $a_j$ vanishes  for some $j\in [1,M]$, while all others $a_{j'}$ remain nonzero, one can derive
 from \eqref{Lam}, \eqref{det} for $m=1,\ldots,j$,
\begin{eqnarray}\label{mum}
\mu_m(a_j)=\begin{cases}
\Lambda_m+O(a_j^2), & {\rm if}\;\;j=1,\ldots,M-1,\\
\Lambda_m+O(a_j), & {\rm if}\;\; j=M, 
\end{cases}\\
\nu_m(a_j)=\begin{cases}
\Lambda_m+O(a_j^2), & {\rm if}\;\;j=1,\ldots,M-1,\\
\Lambda_m+O(a_j), & {\rm if}\;\;j=M. 
\end{cases}.\nonumber
\end{eqnarray}
Repeating  for $j=3,\ldots,M$ the analysis described above leads to representation \eqref{Qa1}.
\end{proof}
Combining \eqref{Qun} and \eqref{Qa1}, one concludes that the polynomial $Q_0(a)$ { on} the right-hand side of \eqref{Qa1}
is uniform, and its degree is zero. Therefore, $Q_0(a)$ does not depend on parameters $a_j$ being just a constant, 
$
Q_0(a)\equiv C_M.
$
To fix this constant, we  choose parameters $\{a_j,b_j\}_{j=1}^M$  according to \eqref{examp}. For such
a choice of the Jacobi matrix, 
the product specified { on} the left-hand side of \eqref{Qa} is given by equation \eqref{pr2}. Comparing the latter
with \eqref{Qa1}, we find
\begin{equation}
C_M=(-1)^{M(M-1)/2} \,\,2^M,
\end{equation}
and arrive { at} equality \eqref{pr0}.

{\remark{
If all $a_j\ne0$, and the  matrix \eqref{HH} satisfying \eqref{even} is 
real, its eigenvalues $\lambda_n$ are also real and non-degenerate.
In this case,  one can order them so  that
\begin{equation}\label{lamord}
\lambda_1<\lambda_2<\ldots<\lambda_N.
\end{equation}
It is possible to show that 
\begin{equation}
 \kappa_n=(-1)^n\, {\rm sign}\, a_M, \quad {\rm for}\quad n=1,\ldots, 2M. 
\end{equation}
 This follows  from the Remark after Theorem 3 in \cite{HOCHSTADT1974435} and Lemma 1.6 in \cite{teschl2000jacobi}.
}
}
\section{Scattering problem for the discrete Schr\"odinger operator \label {SchPr}}
The main issue of this Section is to fix notations for later use.
We briefly summarize some well-known  facts of the 
scattering theory for the discrete Schr\"odinger operator. The  scattering theory for general
 Jacobi operators, which is to some extent
parallel to the scattering theory for the continuous Schr\"odinger equation (see, for example \cite{ChaSab}),
is described in much details in the monograph   \cite{teschl2000jacobi}.

Consider the discrete Schr\"odinger equation  (\ref{eig}) in  $l^2(\mathbb{N})$ 
\begin{eqnarray}\label{Sch}
 \left(\mathbf{H} \psi\right)_j=\lambda\, \psi(j),\\
  \left(\mathbf{H}  \psi\right)_j=    \label{H}
v_j \psi(j)+[2\psi(j)-\psi(j+1)-  \psi(j-1)] , \\
 j=1,2, \ldots,\infty,  \nonumber
\end{eqnarray}
with the potential $v_j\in \mathbb{R}$, supplemented with the Dirichlet boundary condition 
\begin{equation}\label{Dbc}
\psi(0)=0.
\end{equation}
In order to simplify further analysis, only the  potentials $V=\{v_j\}_{j\in \mathbb{N} }$ with a compact support will be considered,
i.e. 
\begin{equation}\label{fr}
v_j=0, \;\; {\rm if}\;\; j> J,
\end{equation}
with some $J\in  \mathbb{N}$. The latter requirement means, that equation \eqref{Sch} looks as the free equation
\begin{equation}
2\psi(j)-\psi(j-1)-\psi(j+1)=\lambda \,\psi(j){, }\label{freeEq}
\end{equation}
{for} $j\ge J+1$.
  
The Jacobi operator   considered in the previous Section reduces to \eqref{H} at $a_j=-1$, and $b_j=2+v_j$.

The  spectrum $\sigma[\mathcal{\mathbf{H} }]$ of the  discrete Schr\"odinger operator   defined by 
\eqref{Sch}-\eqref{fr} consists of the continuous part $\sigma_{cont}[\mathbf{H} ]=[0,4]$ 
and a finite number of  discrete eigenvalues.

At a given $\lambda$, equations \eqref{Sch}, \eqref{H}  with omitted boundary condition \eqref{Dbc}
 have two linearly  independent solutions, and the general 
solution of  \eqref{Sch}, \eqref{H}  can be written as their linear combination. 
Following \cite{teschl2000jacobi}, we introduce the Wronskian of two sequences 
$\{\psi_1(j)\}_{j\in \mathbb{N} }$, and 
$\{\psi_2(j)\}_{j\in \mathbb{N} }$,
\begin{equation}
W_j(\psi_1,\psi_2)=-[\psi_1(j)\psi_2(j+1)-\psi_1(j+1)\psi_2(j)], \quad j=0,1,2,\ldots
\end{equation}
It  does not depend on $j$, if $\psi_1(j)$ {and} $\psi_2(j)$  are solutions of equations \eqref{Sch}, \eqref{H}.

Let us turn now to the scattering problem associated with equations \eqref{Sch}-\eqref{fr}, which 
corresponds to  the case $0<\lambda<4$.
Instead of the parameter $\lambda\in \sigma_{cont}[\mathbf{H} ]$, it is  also convenient  to  use the  momentum $p$ and the related complex
parameter $z=e^{i p}$:
\begin{equation}\label{plam}
\lambda=2-2 \cos p=2-z-z^{-1}.
\end{equation}
Three solutions of  \eqref{Sch} are important for the scattering problem.
\begin{itemize}
\item
The {\it fundamental solution} $\varphi(j,p)$, which is fixed by the boundary conditions
\begin{equation}\label{bc8}
\varphi(0,p)=0, \quad \varphi(1,p)=1.
\end{equation}
\item
Two {\it Jost solutions} $f(j,p)$, and $f(j,-p)$, which are determined by their behavior at large $j$,
and describe the out- and in-waves, respectively,
\begin{equation}\label{Jas}
f(j,\pm p)=  \exp(\pm i   p j)=z^{\pm j}, \quad{\rm {at}}\quad  j\ge J+1.
\end{equation}
{Note, that this formula is exact only in the  considered case of the potential $V$ with a compact support. In the general case, it must
by replaced by the asymptotical formula $f(j,\pm p)\sim  \exp(\pm i   p j)=z^{\pm j}$, as $j\to\infty$. This comment also relates to equations
 \eqref{phieta} and \eqref{spcon} below.}
\end{itemize}
The fundamental solution $\varphi(j,p)$ can be represented  as a linear combination of two Jost
solutions,
\begin{equation}\label{FJ}
\varphi(j,p)=\frac{i }{2\sin p}\left[F(p)f(j,-p)-F(-p)f(j,p)\right], \quad 0<p<\pi.
\end{equation}
The complex coefficient $F(p)$ in the above equation is known as  the Jost function.
It is determined by \eqref{FJ} for 
 real momenta $p$ in the 
 interval $p\in (-\pi,\pi)$, where it satisfies the relation
 \begin{equation}
 F(-p)=[F(p)]^*,
 \end{equation}
and  can be written as
\begin{equation}\label{Fs}
F(p)=\exp[{\sigma(p)- i  \eta(p) }],
\end{equation}
with real $\sigma(p)$ and $\eta(p)$.
At large $j\ge J+1$, the fundamental solution behaves as
\begin{equation}\label{phieta}
\varphi(j,p)=\frac{A(p)}{\sin p} \sin[p\, j+\eta(p)],
\end{equation}
where  $A(p)=\exp [\sigma (p)]$ is the scattering amplitude, and $\eta(p)$ is the scattering phase.
The latter can be defined in such a way, that $\eta(-p)=-\eta(p)$ for $-\pi<p<\pi$.

Note, that the Jost function $F(p)$ can be also represented as the Wronskian of the Jost and fundamental solutions
\begin{equation}
F(p)=W_j(\varphi(p),f(p)), 
\end{equation}
with arbitrary $j\in \mathbb{N}$.

The following exact representation holds for the Jost function $F(p)$ in terms of the fundamental solution $\varphi(j,p)$:
\begin{equation} \label{JF}
F(p)=1+\sum_{j=1}^J e^{i  pj}v_j \varphi(j,p),
\end{equation}
cf. equation (1.4.4) in \cite{ChaSab} in the continuous case.

For $|z|=1$, denote by $\hat{F}(z)$ the Jost function $F(p)$ expressed in { terms} of 
the complex parameter $z$: $F(p)=\hat{F}(z=e^{i  p})$.
The function  $\hat{F}(z)$ can be analytically continued into the circle $|z|<1$, where it has finite number of 
zeros $\{\alpha_n\}_{n=1}^\mathfrak{N}$. These zeroes determine the discrete spectrum $\{\lambda_n\}_{n=1}^\mathfrak{N}$  of the
problem \eqref{Sch}-\eqref{Dbc}:
\begin{equation}
\lambda_n=2-\alpha_n-\alpha_n^{-1}, \quad {\rm for  } \quad n=1,\ldots,\mathfrak{N}.
\end{equation}
Of course, these eigenvalues are real in the boundary problem with a real potential. 

Besides \eqref{fr}, two further restrictions will be imposed in the sequel on the 
 potential $V$.
\begin{itemize}
\item[(i)] \label{Jzl1}
The Jost function $\hat{F}(z)$ associated with such a potential should not have zeroes inside the circle $|z|<1$, i.e. 
$\mathfrak{N}=0$. In other words, 
the spectrum $\sigma[\mathbf{H} ]$ should be purely continuous. 
\item[(ii)] \label{Jzpm1}
The Jost function $\hat{F}(z)$ should take non-zero values at $z=\pm 1$: $\hat{F}(1)\ne 0$, and $\hat{F}(-1)\ne 0$.
\end{itemize}
Conditions (i) and (ii) imply, that the operator $\mathbf{H}$ does not have bound and semi-bound states 
\cite{ChaSab}, 
respectively.  

Since $ \varphi(j,p)$ is a polynomial of the spectral parameter $\lambda=2-z-z^{-1}$  of the degree $j-1$, 
the Jost function \eqref{JF} expressed in the parameter $z$ is a polynomial of the degree $2J-1$:
\begin{equation}\label{Jz}
\hat{F}(z)=1+\sum_{j=1}^{2J-1}c_j(V)\,z^j=\prod_{n=1}^{2J-1}\left[1-\frac{z}{z_n(V)}\right],
\end{equation}
where the { real} coefficients $c_j(V)$ polynomially depend on the potential $v_j$, $j=1,\ldots,J$. 
This indicates, that the set of Jost polynomial \eqref{Jz} has the  co-dimension $J-1$ in the
$2J$-dimensional linear space of polynomials of  the degree $2J-1$.
Due to the constrains (i), (ii), we get
\begin{equation}\label{a}
|z_n(V)|>1, \quad{\rm for\; all} \quad n=1,\ldots, 2J-1.
\end{equation}

Let us periodically continue the scattering phase $\eta(p)$ from the interval $(-\pi,\pi)$ to the whole real axis
$p\in \mathbb{R}$.
It follows from \eqref{a}, that for a potential $V$ satisfying \eqref{fr}, (i), (ii), the scattering phase 
 is a {smooth} odd $2\pi$-periodical function in the whole real axis:
 $\eta(p)\in C^\infty(\mathbb{R}/2\pi \mathbb{Z})$.
 
The notation $\delta(\lambda)$  will be used for the scattering phase $\eta(p)$ expressed in terms of the 
spectral parameter
$\lambda$: $\eta(p)=\delta(\lambda=2-2 \cos p)$, for $0\le\lambda\le4$, and $0\le p\le \pi$. Conditions 
(i), (ii)  guarantee, that 
\begin{equation}\label{del}
\delta(0)=\delta(4)=0.
\end{equation}
\section{Identities on the scattering data for the discrete \\Schr\"odinger  operator \label{infM}}
In this Section we apply Theorem  \ref{Th2} to study some spectral properties of the
discrete Schr\"odinger operator acting in ${l}^2( \mathbb{N})$, and prove the following 
\begin{theorem} \label{Th3} The scattering phase $\delta(\lambda)$ for the 
 discrete Schr\"odinger operator  $\mathbf{H}\in {\rm End} [{l}^2( \mathbb{N})]$ which is defined by  \eqref{H} with the Dirichlet boundary condition \eqref{Dbc}, and arbitrary potential 
 $V=\{v_j\}_{j=1}^\infty$ { obeying} \eqref{fr}, (i), and (ii),
must satisfy the equality 
\begin{equation}\label{EQ}
\int_0^4 d\lambda\,\delta(\lambda)\,
\frac{\lambda-2}{\lambda(\lambda-4)}+\frac{1}{\pi}\int_0^4 d\lambda_1\, \delta(\lambda_1)\,\mathcal{P}\!\!
\,\int_0^4 d\lambda_2\, \frac{\delta'(\lambda_2)}{\lambda_2-\lambda_1}=0,
\end{equation}
where $\mathcal{P}\!\!\int$ indicates the principal value integral. 

Equality \eqref{EQ} can be rewritten as well in 
terms of the  Jost function $\hat{F}(z)$:
\begin{equation}\label{Jost}
{{\log}\, \hat{F}(z=1)+{\log} \,\hat{F}(z=-1)}+\oint_{|z|=1} \frac{d z}{2\pi  i }\, {\log}\, [\hat{F}(1/z)]\,\frac{d\,{\log}[\hat{F}(z)]}{d z}=0,
\end{equation}
where the integration path  {  is counter-clockwise oriented}. 

Equality \eqref{Jost} can be also written in 
terms the zeros $\{z_n(V)\}_{n=1}^{2J-1}$ of the  Jost function $\hat{F}(z)$:
\begin{equation}\label{Jzer}
\prod_{1\le m<n\le 2J-1}\left[1-\frac{1}{z_n(V)z_m(V)}\right]=1.
\end{equation}
\end{theorem} 

\begin{proof}
To prove \eqref{EQ}, let us consider the discrete Schr\"odinger  eigenvalue problem  \eqref{eig}-\eqref{BC}
in the finite interval $1\le j \le N=2M$, with $a_j=-1$,  $M>J$, and with the 
even potential $V^{(M)}=\{v_j^{(M)}=b_j-2\}_{j=1}^{2M}$, which restriction to the interval 
$[1,M]$ coincides with that of the potential $V=\{v_j\}_{j=1}^\infty$: 
\begin{equation}\label{potM}
v_j^{(M)}=\begin{cases}
 v_j, \quad{\rm if} \quad j\le J,\\
  0, \quad{\rm if} \quad J<j\le 2M-J,\\
 v_{2M+1-j}, \quad{\rm if} \quad 2M-J< j \le 2M .
\end{cases}
\end{equation}
As it was explained in the Introduction,  the eigenfunctions $\Psi_l(j)$ of such a problem
are either even, or odd with respect to the reflection \eqref{refl}, see equations \eqref{eigv}.
In both cases, the equality 
\begin{equation}\label{psiM}
[\Psi_l(M)]^2=[\Psi_l(M+1)]^2
\end{equation}
must hold. On the other hand, the eigenfunction $\Psi_l(j)$ corresponding to the eigenvalue 
$\lambda_l$ obeys the Dirichlet boundary condition at $j=0$, and, therefore, is proportional to 
the fundamental solution $\varphi(j,p_l)$, 
\begin{equation}\label{psij}
\Psi_l(j)=C_l \,\varphi(j,p_l),
\end{equation}
where $p_l$ and $\lambda_l$ are related according
to \eqref{plam}. Since the potential $v_j$ is zero at the sites $j=M-2,M-1,M,M+1,M+2$
near the middle of the interval $[1,2M]$, the fundamental solution $\varphi(j,p_l)$ has the 
form \eqref{phieta} at $j=M,M+1$. Substitution of \eqref{phieta} and \eqref{psij} into 
\eqref{psiM} leads to the equation
\begin{equation}\label{spcon}
\sin^2[p_l M+\eta(p_l)]=\sin^2[p_l (M+1)+\eta(p_l)],
\end{equation}
which determines the spectrum $\lambda_l\equiv\omega(p_l)$ in 
terms of the scattering phase $\eta(p)$. The equivalent, but more 
 convenient form of equation
\eqref{spcon} reads
\begin{equation}
(2M+1)\,p_l +2\,\eta(p_l)=(2M+1)\,k_l, \label{pk}
\end{equation}
where  $k_l$ were defined in \eqref{kn}.

For an arbitrary $M>J$, let us put $a_j=-1$ for all $j\in[0,M]$ in equality \eqref{pr0}, and take the logarithm of  the ratio of two versions of this equality written for the potential \eqref{potM}, and for 
the zero potential, respectively. 
As { a} result, we get
\begin{equation}\label{Sm0}
\sum_{m=1}^M S_m=0,
\end{equation}
where
\begin{equation}\label{Sm}
S_m=\sum_{n=1}^M \left[
\ln|\omega(p_{2n-1})-\omega(p_{2m})|-\ln|\omega(k_{2n-1})-\omega(k_{2m})|
\right].
\end{equation}
{
In order to proceed to the large-$M$ limit { on} the right-hand side, let us introduce three one-parametric sets of the $C^\infty$-functions $k(p;\epsilon)$, $p(k;\epsilon)$, and 
$F(k,k';\epsilon)$, where $\epsilon$ is a small parameter lying in the interval 
 $0\le\epsilon\le \delta$, and 
$\delta$ is some fixed  positive number such that
\begin{equation}
 \delta<\frac{1}{2}\max_{p\in {\mathbb R}} |\eta'(p)|^{-1}.
\end{equation}
The function $k(p;\epsilon)$ from the first set is defined in the real axis $p\in{\mathbb R}$ by the equation 
\begin{equation}
  k(p;\epsilon)=p+2 \,\epsilon\, \eta(p).
\end{equation}
At a fixed $\epsilon\in[0,\delta]$, the function 
$k(p;\epsilon)$  monotonically increases in $p$ being  odd  and quasiperiodic,  
\[
\partial_p k(p;\epsilon)>0,\quad k(-p;\epsilon)=-k(p,\epsilon),\quad  k(p+2\pi;\epsilon)=k(p;\epsilon)+2\pi.
\]

For $\epsilon\in[0,\delta]$, the inverse function $p(k;\epsilon)$ is a well defined odd quasiperiodic
$C^\infty$-function of $k\in {\mathbb R}$, which also smoothly depends on the parameter
$\epsilon$. The Taylor expansion of $p(k,\epsilon)$ in $\epsilon$ reads as
\begin{equation}
p(k;\epsilon)= k-\epsilon\, \eta(k)+4 \epsilon^2\eta(k)\,\eta'(k)+\sum_{j=3}^\infty  \epsilon^j A_j(k),
\end{equation}
where $A_j(k)$ are $2\pi$-periodic odd bounded functions of $k$.

The function representing  the third set, 
\begin{equation}
F(k,k';\epsilon)=\ln \frac{\omega[p(k;\epsilon)]-\omega[p(k';\epsilon)]}{\omega(k)-\omega(k')}
\end{equation}
is $2\pi$-periodic and even in both arguments $k,k'\in  {\mathbb R}$,
and $F(k,k';\epsilon)\in C^\infty\left[ {\mathbb R}\times  {\mathbb R}\times [0,\delta]\right]$.
Its Taylor expansion in $\epsilon$
\begin{equation}\label{TaylF}
F(k,k';\epsilon)=\sum_{n=1}^\infty \frac{\epsilon^n}{n!} \,
{\partial_\epsilon^n F(k,k';\epsilon)}\Big|_{\epsilon=0}
\end{equation}
converges  at $0\le\epsilon\le\delta$  uniformly in $\langle k,k'\rangle\in {\mathbb R}^2$.
The two initial partial derivatives in this expansion read as
\begin{align}\label{d1F}
&{\partial_\epsilon F(k,k';\epsilon)}\Big|_{\epsilon=0}=
2\,\frac{\omega'(k')\,\eta(k')-\omega'(k)\,\eta(k)}
{\omega(k)-\omega(k')},\\
&{\partial_\epsilon^2 F(k,k';\epsilon)}\Big|_{\epsilon=0}=
4\,\Bigg\{
 \frac{\left[\omega'(k)\,\eta^2(k)\right]'-\left[\omega'(k')\,\eta^2(k')\right]'}
{\omega(k)-\omega(k')}
-\left[ \frac{\omega'(k')\,\eta(k')-\omega'(k)\,\eta(k)}
{\omega(k)-\omega(k')}
\right]^2\Bigg\}.
\end{align}
These and all higher derivatives 
$\frac{\partial^n F(k,k';\epsilon)}{\partial \epsilon^n}\Big|_{\epsilon=0}$, $n=3,\ldots$, in the Taylor expansion 
\eqref{TaylF} are the $2\pi$-periodic  $C^\infty[{\mathbb R}^2]$-functions of their arguments $\langle k,k'\rangle \in {\mathbb R}^2$, which are therefore bounded in ${\mathbb R}^2$ together with all their 
partial derivatives in $k$ and $k'$.

Let us now rewrite the sum \eqref{Sm} as
\begin{equation}
S_m=\sum_{n=1}^M F(k_{2n-1},k_{2m};\epsilon)\big|_{\epsilon=(2M+1)^{-1}},
\end{equation}
and expand the right-hand side in $\epsilon$ to the second order,
\begin{eqnarray}\label{Sm3}
&&S_m=\sum_{n=1}^M 
\Bigg[\frac{\partial_\epsilon F(k_{2n-1},k_{2m};\epsilon)\big|_{\epsilon=0}}{2M+1}
+\frac{\partial_\epsilon^2 F(k_{2n-1},k_{2m};\epsilon)\big|_{\epsilon=0}}{2(2M+1)^2}\\
&&+\delta_{3}\,F(k_{2n-1},k_{2m};\epsilon)\big|_{\epsilon=(2M+1)^{-1}}\Bigg].\nonumber
\end{eqnarray} 
It follows from the above analysis, 
that the correction term in the square brackets is uniformly bounded in $n$ and $m$, 
\begin{equation}
|\delta_{3}\,F(k_{2n-1},k_{2m};\epsilon)|<\epsilon^3 C, 
\end{equation}
with some positive $C$ independent on $n,m$. Taking this into account, one obtains from 
\eqref{d1F}-\eqref{Sm3},
\begin{equation}\label{SMA}
S_m=S_m^{(0)}+ S_m^{(1)}+O(M^{-2}),
\end{equation}
where
\begin{eqnarray}\label{S0}
&&S_m^{(0)}=\frac{1}{2M+1}\sum_{n=1}^M \partial_\epsilon F(k_{2n-1},k_{2m};\epsilon)\big|_{\epsilon=0},\\
&&S_m^{(1)}=\frac{X_m}{\pi(2M+1)}.\label{S1a}
\end{eqnarray}
where
\begin{eqnarray}\label{Xmsum}
&&X_m=\frac{ \pi}{2(2M+1)}\,
\sum_{n=1}^M \partial_\epsilon^2 F(k_{2n-1},k_{2m};\epsilon)\big|_{\epsilon=0}, 
\end{eqnarray}
and the $m$-dependent correction term $O(M^{-2})$ is again uniformly bounded in $m$. 
Replacing  in \eqref{Xmsum}  the Riemann sum of the regular function of  the momenta $k_{2n-1}$ by the
corresponding integral in the limit of large $M$, one obtains 

\begin{equation}
\label{X}  
X_m=\frac{1}{4}\,\int_0^\pi d q\,\partial_\epsilon^2 
F(q,k_{2m};\epsilon)\big|_{\epsilon=0}+O(M^{-1}).
\end{equation}
The $m$-dependent correction term $O(M^{-1})$ is uniformly bounded in $m$, 
since \newline $\partial_\epsilon^2 
F(q,k_{2m};\epsilon)\big|_{\epsilon=0}\in C^\infty [{\mathbb R}/(2\pi{\mathbb Z})\times {\mathbb R}/(2\pi {\mathbb Z})]$.

Calculation of the large-$M$ asymptotics of $S_m^{(0)}$ is more delicate. 
First, we extend summation in \eqref{S0} in the index $n$ from 1 till $2M+1$,
\begin{eqnarray}\nonumber
S_m^{(0)}=\frac{2}{2M+1}\sum_{n=1}^{M}R_m(k_{2n-1})=\\
\frac{2}{2M+1}\left[-
\frac{R_m(k_{2M+1})}{2}+\frac{1}{2}\sum_{n=1}^{2M+1} R_m(k_{2n-1})\right],\label{S0b}
\end{eqnarray}
where 
\begin{equation}\label{Rm}
R_m(q)=\frac{1}{2}{\partial_\epsilon F(q,k_{2m};\epsilon)}\Big|_{\epsilon=0}=\frac{\omega'(k_{2m})\,\eta(k_{2m})-\omega'(q)\,\eta(q)}
{\omega(q)-\omega(k_{2m})}.
\end{equation}
In \eqref{S0b} we have taken into account the reflection symmetry  $R_m(q)=R_m(2\pi-q)$ 
of the function \eqref{Rm}, which leads to
$R_m(k_{2n-1})=R_m(k_{2(2M+1-n)+1})$.

Since $k_{2M+1}=\pi$, and $\eta(\pi)=0$, $\omega(\pi)=4$, we  get from \eqref{Rm}
\begin{equation}\label{Rm1}
R_m(k_{2M+1})=\frac{\omega'(k_{2m})\,\eta(k_{2m})}
{4-\omega(k_{2m})}.
\end{equation}
The sum  in the second line of \eqref{S0b} reads as 
\begin{equation}\label{Rm2}
\sum_{n=1}^{2M+1} R_m(k_{2n-1})=\sum_{n=1}^{2M+1} R_m\left(2\pi\,\frac{ n-1/2}{2M+1}\right).
\end{equation}
Since $R_m(q)\in C^\infty(\mathbb{R}/2\pi \mathbb{Z})$, 
this sum can be  replaced 
 with exponential accuracy by the integral at large $M\to\infty$:
\begin{equation}\label{sumR}
\sum_{n=1}^{2M+1} R_m\left(2\pi\,\frac{ n-1/2}{2M+1}\right)=\frac{2M+1  }{2\pi}
\int_0^{2\pi}d q \,R_m(q)+ o(M^{-\mu}),
\end{equation}
where $\mu$ is an arbitrary positive number, see formula 25.4.3 in \cite{AbrSt}.
Furthermore, since 
${\partial_\epsilon F(q,k;\epsilon)}\Big|_{\epsilon=0}\in C^\infty [{\mathbb R}/(2\pi{\mathbb Z})\times {\mathbb R}/(2\pi {\mathbb Z})]$, the $m$-dependent correction term $o(M^{-\mu})$ is uniformly bounded 
in $m$.
}

Taking into account \eqref{Rm}, 
the integral { on} the right-hand side { of \eqref{sumR}} can be written as
\begin{eqnarray}\label{Rint}
&&\int_0^{2\pi}d q \,R_m(q)=\omega'(k_{2m})\,\eta(k_{2m}) \,\mathcal{P} \!\int_0^{2\pi}\frac{d q }{\omega(q)-\omega(k_{2m})}\\
&&-\mathcal{P} \!\int_0^{2\pi} d q\,\frac{\omega'(q)\,\eta(q)}{\omega(q)-\omega(k_{2m})}.\nonumber
\end{eqnarray}
The principal value integral in the first term { on} the right-hand side vanishes
\begin{equation}
\mathcal{P} \! \int_0^{2\pi}\frac{d q }{\omega(q)-\omega(k_{2m})}=
2\, \mathcal{P} \!\int_0^{\pi}\frac{d q }{\omega(q)-\omega(k_{2m})}=0
\end{equation}
due to the equality
\begin{eqnarray}\nonumber
 &&\mathcal{P}\!\int_0^\pi\, \frac{ d q	}{[\omega(q)-\omega(k)]^\nu}\equiv \\
&&\frac{1}{2}\lim_{\epsilon\to +0}\left\{
\int_0^\pi\, \frac{ d q	}{[\omega(q+i \epsilon)-\omega(k))]^\nu}+\int_0^\pi\, \frac{ d q	}{[\omega(q-i   \epsilon)-\omega(k)]^\nu}
\right\}=0,\label{eqP}
\end{eqnarray}
which can be easily proved for $\omega(p)=2-2 \cos p$, all  $\nu\in \mathbb{N}$, and $0<k<\pi$. 
The integral in the second line of \eqref{Rint} reduces after { the  change  $ \lambda=\omega(q)$  of the integration variable} 
to the form 
\[
\mathcal{P} \!\int_0^{2\pi} d q\,\frac{\omega'(q)\,\eta(q)}{\omega(q)-\omega(k_{2m})}=2\,I[\omega(k_{2m})],
\]
where \begin{equation}\label{IL}
I(\Lambda)=\mathcal{P} \!\int_0^{4}
d\lambda\,\frac{\delta(\lambda) }{\lambda-\Lambda}, \quad{\rm with}\quad 0<\Lambda<4.
\end{equation}

Collecting \eqref{S0b}-\eqref{IL}, we get 
\begin{equation}\label{Sm0a}
S_m^{(0)}=-\frac{I[\omega(k_{2m})]}{\pi}-\frac{1}{2M+1}\frac{\omega'(k_{2m})\,\eta(k_{2m})}
{4-\omega(k_{2m})}+{ o(M^{-\mu}),}
\end{equation}
where the $m$-dependent error term $o(M^{-\mu})$ is uniformly bounded in $m$.
Thus, we obtain from \eqref{Sm0} and \eqref{SMA}  the equality
\begin{equation}\label{limS}
\lim_{M\to\infty}\sum_{m=1}^M S_m^{(0)}+\lim_{M\to\infty}\sum_{m=1}^MS_m^{(1)}=0,
\end{equation}
where $S_m^{(0)}$ and $S_m^{(1)}$ are given by equations \eqref{Sm0a}, and \eqref{S1a}, \eqref{X}, respectively. 

In the second term, we can replace  with sufficient accuracy  the summation in $m$ by integration in the momentum $k$:
\begin{eqnarray}\nonumber
\sum_{m=1}^M S_m^{(1)}=\frac{1}{2\pi^2}\int_0^\pi d k\int_0^\pi d q\,
 \frac{\left[\omega'(q)\,\eta^2(q)\right]'-\left[\omega'(k)\,\eta^2(k)\right]'}
{\omega(q)-\omega(k)}\\
-\frac{1}{2\pi^2}\int_0^\pi d k\int_0^\pi d q\,\left[ \frac{\omega'(k)\,\eta(k)-\omega'(q)\,\eta(q)}
{\omega(q)-\omega(k)}
\right]^2+O(M^{-1}).\label{S1b}
\end{eqnarray}
The first integral { on} the right-hand side vanishes due to equality \eqref{eqP}
with $0<k<\pi$, and $\nu=1$. The second line in \eqref{S1b} can be transformed as follows
\begin{eqnarray}\label{aux}
&&-\frac{1}{2\pi^2}\int_0^\pi d k\int_0^\pi d q\,\left[ \frac{\omega'(k)\,\eta(k)-\omega'(q)\,\eta(q)}
{\omega(q)-\omega(k)}
\right]^2= \\
&& -\frac{1}{\pi^2}\int_0^\pi \!d k\, [\omega'(k)\,\eta(k)]^2\,\mathcal{P}\!\!\int_0^\pi  \,\frac{d q }
{[\omega(q)-\omega(k)]^2}+\nonumber \\
&&
\frac{1}{\pi^2}\int_0^\pi \!d k\,\mathcal{P}\!\!\int_0^\pi\!d q \,\frac{  \omega'(k)\,\eta(k) \omega'(q)\,\eta(q)}
{[\omega(q)-\omega(k)]^2}.\nonumber
\end{eqnarray} 
The first term { on} the right-hand side vanishes due to equality \eqref{eqP} with $\nu=2$. 
Then, substituting for $\lambda_1=\omega(k)$, $\lambda_2=\omega(q)$, and integrating by parts, we obtain from the second term { on} the right-hand side of \eqref{aux}
\begin{equation}\label{limS1}
\lim_{M\to\infty}\sum_{m=1}^MS_m^{(1)}=\frac{1}{\pi^2}\int_0^4 d\lambda_1\, \delta(\lambda_1)\,\mathcal{P}\!\!
\,\int_0^4 d\lambda_2\, \frac{\delta'(\lambda_2)}{\lambda_2-\lambda_1}.
\end{equation}

Let us turn  now to calculation of the first term { on} the left-hand side of equality \eqref{limS}. At large $M$, one obtains
\begin{eqnarray}\label{Sma}
 \sum_{m=1}^M S_m^{(0)} =-\frac{1}{2\pi}
\int_0^{\pi} d k\,\frac{\omega'(k)\,\eta(k)}{4-\omega(k)}-\frac{1}{\pi}\sum_{m=1}^M I[\omega(k_{2m})]+O(M^{-1}).
\end{eqnarray}
The first term { on} the right-hand side equals to $I(4)/(2\pi)$.
The large-$M$ asymptotics of the sum { on} the right-hand side can be found as follows
\begin{eqnarray}\nonumber
\sum_{m=1}^M I[\omega(k_{2m})]=
-\frac{I(0)}{2}+\frac{1}{2}\sum_{m=1}^{2M+1} I[\omega(k_{2m})]=\\
 -\frac{I(0)}{2}+\frac{2M+1}{4\pi}\int_0^{2\pi} d k\,I[\omega(k)]+O(M^{-\mu}),\label{S0d}
\end{eqnarray}
where $\mu$ is an arbitrary positive number. In deriving \eqref{S0d} we have taken 
into account, that $I[(\omega(k)]$ is the $2\pi$-periodical function of $p$ in $\mathbb{R}$, which is continuous 
with 
all its derivatives  [i.e.,  $I[(\omega(k)]\in C^\infty(\mathbb{R}/2\pi \mathbb{Z})$], and formula 25.4.3 in \cite{AbrSt}.
The integral { on} the right-hand side vanishes due to equality \eqref{eqP} with $\nu=1$:
\begin{equation}\label{Iint}
\int_0^{2\pi} d k\,I[\omega(k)]=\int_0^{2\pi} d k\,\mathcal{P}\!\!\int_0^4 d\lambda\, \frac{\delta(\lambda)}{\lambda-\omega(k)}=
\int_0^4 d\lambda\, \delta(\lambda)\,\mathcal{P}\!\!\int_0^{2\pi} \frac{d k}{\lambda-\omega(k)}=0.
\end{equation}
Collecting \eqref{Sma}-\eqref{Iint}, we come to the simple formula
\begin{equation}\label{limS0}
\lim_{M\to\infty}\sum_{m=1}^M S_m^{(0)} =\frac{I(0)+I(4)}{2\pi}.
\end{equation}
From \eqref{IL}, \eqref{limS}, \eqref{limS1}, \eqref{limS0}, we arrive { at} the final result \eqref{EQ}.

Derivation of \eqref{Jost} from \eqref{EQ} is straightforward being based on \eqref{Fs}, and exploits 
properties (i), (ii), which
were supposed to hold for the 
Jost function $\hat{F}(z)$ in Section \ref{SchPr}.

The integrand of the integral { on the }  left-hand side of equation \eqref{Jost} is meromorphic in the region $|z|>1$ 
having there only simple poles at $z=z_n$, $n=1,\ldots, 2J-1$ with residues $\ln \hat{F}(1/z_n)$.  
After straightforward application of the Cauchy's residue theorem, 
one finds from \eqref{Jost},
\begin{equation}\label{Jost1}
\ln \hat{F}(1)+\ln \hat{F}(-1)-\sum_{n=1}^{2J-1}\ln \hat{F}(1/z_n)=0.
\end{equation}
{
Substitution  of the right-hand side of \eqref{Jz} into this relation yields,
\begin{equation*}
\left[\prod_{1\le m<n\le 2J-1}\left(1-\frac{1}{z_n z_m}\right)\right]^2=1.
\end{equation*}
In order prove equality \eqref{Jzer}, it remains to show that the product in the 
square brackets on the  left-hand side  is positive. The latter statement follows immediately from the two properties of 
the zeroes  $z_n$ of the  Jost  function $\hat{F}(z)$.
\begin{enumerate}
\item $|z_n|>1$ for all $n=1,\ldots,2J-1$, see equation \eqref{a}.
\item The numbers $z_n$ 
are either real,  or else occur as complex conjugate pares in the set $\{z_n\}_{n=1}^{2J-1}$ 
of roots of the real polynomial \eqref{Jz}.
\end{enumerate}}
\end{proof}
\begin{example}
Let us check directly that equality \eqref{Jzer} holds for  {$J=2$}. 
In this case, the Jost function \eqref{Jz} is the polynomial of the third degree,
\begin{equation}\label{Jost2}
\hat{F}(z)=1+z(v_1+v_2)+z^2 v_1v_2+z^3v_2,
\end{equation}
which is parametrized by two parameters $v_1,v_2$.
Let us rewrite equation \eqref{Jzer} in the form 
\begin{equation}\label{symeq}
\frac{\sigma_3^2-\sigma_3\sigma_1+\sigma_2-1}{\sigma_3^2}=1
\end{equation}
in terms of the symmetric polynomials of the Jost function zeroes,
\begin{eqnarray*}
&&\sigma_1=z_1+z_2+z_3,\\
&&\sigma_2=z_1z_2+z_1z_3+z_2z_3,\\
&&\sigma_3=z_1z_2z_3.
\end{eqnarray*}
Exploiting the Vieta's theorem, one obtains from  \eqref{Jost2},
\begin{eqnarray*}
&&\sigma_1=-v_1,\\
&&\sigma_2=\frac{v_1+v_2}{v_2},\\
&&\sigma_3=-\frac{1}{v_2}.
\end{eqnarray*}
Upon substitution of the right-hand sides into \eqref{symeq}, we convince ourselves 
that  equality \eqref{symeq} holds  for arbritrary complex $v_1$ and $v_2$. 
\end{example}
\section{Conclusions and open problems\label{conc}}
We have studied some  spectral properties of two classes of Jacobi operators.

The first class is represented by the real $2M$-dimensional { Jacobi matrices with the refection symmetry
\eqref{even}}. We have  proved 
in Theorem \ref{Th2} a new polynomial identity \eqref{pr0}, which relates the eigenvalues 
 of such Jacobi matrices with their matrix elements.

The second class of Jacobi operators considered in the present paper is represented by the discrete 
Schr\"odinger operators acting in $l^2(\mathbb{N})$ with purely continuous spectrum, 
which potentials have a compact support. 
Note, that the infinite-dimensional Jacobi matrix corresponding to the discrete
Schr\"odinger operator has the 
super- and sub-diagonal matrix elements $a_j=-1$ for all $j\in \mathbb{N}$.
The scattering data of such operators must satisfy three identities \eqref{EQ}, 
\eqref{Jost}, and \eqref{Jzer}
which are proved in Theorem~\ref{Th3}.

Obtained result could be extended and generalized in several directions.

We did not try to prove identities \eqref{EQ}, 
\eqref{Jost} 
 for the most general case of the discrete semi-infinite scattering problem. 
It is natural to expect, however, that
they should hold for some more general class of potentials, which  vanish fast enough at infinity, 
though do not have a compact support. On the other hand, in the case of the potentials with non-empty discrete spectrum
and/or semi-bound states (the latter appear if the Jost function has zeroes at the end points of the 
continuous spectrum, see \cite{ChaSab}), some modified forms of \eqref{EQ}, \eqref{Jost}  should also exist.

It should be noted also, that  no counterparts of Theorems \ref{Th2} and \ref{Th3} 
are known in the continuous Sturm-Liouville theory. Simple heuristic arguments exploiting the formal continuous 
limit of the discrete boundary problem \eqref{eig}-\eqref{BC} and identity \eqref{pr0} lead to the following tentative 
form of the extension of Theorem \ref{Th2} to the continuous case.

{\bf Hypothesis.} {\it Let $\mathcal{H}^{(\pm)}$ be two Schr\"odinger operators 
\begin{equation}\label{Scro}
\mathcal{H}^{(\pm)}=-\partial_z^2\pm v(z)
\end{equation}
acting in "the appropriate space of function" defined in the interval $0\le z\le \pi$, which satisfy the  Dirichlet boundary conditions $\psi(0)=\psi(\pi)=0$. Let the  potential $v(z)$ in \eqref{Scro} be a  
 real even function,
\[
v(\pi-z)=v(z),
\] 
which belongs to some space $\mathcal L$ of "good enough" potentials.
The space $\mathcal L$ should contain the   { piecewise continuous potentials defined on  $[0,\pi]$}.

Denote by $\{\mu_m^{(\pm)},\nu_m^{(\pm)}\}_{m\in\mathbb{N}}$ the naturally ordered 
eigenvalues of the operators $\mathcal{H}^{(\pm)}$, 
which correspond to the eigenfunctions with different parities:
\begin{eqnarray}
&&\mathcal{H}^{(\pm)} \,\psi_{m}^{(\pm,ev)}= \mu_m^{(\pm)}\,\psi_{m}^{(\pm,ev)}, 
\quad \psi_{m}^{(\pm,ev)}(\pi-z)=\psi_{m}^{(\pm,ev)}(z),\\\nonumber
&&\mathcal{H}^{(\pm)} \,\psi_{m}^{(\pm,od)}= \nu_m^{(\pm)}\,\psi_{m}^{(\pm,od)}, 
\quad \psi_{m}^{(\pm,od)}(\pi-z)=-\psi_{m}^{(\pm,od)}(z),\\
&&\mu_m^{(\pm)}<\mu_{m+1}^{(\pm)}, \quad \nu_m^{(\pm)}<\nu_{m+1}^{(\pm)}, \quad
m=1,\ldots,\infty.\nonumber
\end{eqnarray}
Then the following identity should hold
\begin{equation}\label{pr3}
\prod_{m=1}^\infty\prod_{n=1}^\infty\frac{\mu_m^{(+)}-\nu_{n}^{(+)}}{(2 m)^2-(2n-1)^2}=
\prod_{m=1}^\infty\prod_{n=1}^\infty\frac{(2 m)^2-(2n-1)^2}{\mu_m^{(-)}-\nu_{n}^{(-)}}.
\end{equation}
}

We have performed numerical check of formula \eqref{pr3} for the probe singular
potential 
$v(z)= A\, \delta(z-\pi/2)$ with several values of the parameter $A$ in the interval $1\le A\le 5$ by  
cutting the infinite products in $n$ and $m$ in \eqref{pr3} at different values of $n_{max}=m_{max}=N$, 
with $10\le N\le100$.
The results of our numerical calculations agree well with \eqref{pr3}. 

It would be interesting to find the precise form and the direct proof of the above statement. 

\section*{Funding Statement} 
This work was funded  by Deutsche Forschungsgemeinschaft (DFG),  via grant Ru 1506/1.
\section*{Acknowledgement}
I am thankful to H.~W.~Diehl for many fruitful discussions. 
\section*{Appendix. Proof of equality \eqref{pr2} \label{free}}
Let us start from the following auxiliary
\begin{lemma} The following equality holds for all natural  $M$ and integer  $n$:
\begin{equation}\label{l2}
\prod_{m=1}^{2M+1}\left\{4 \sin^2\left[\frac{ (2m-1)\pi}{2(2M+1)}-\alpha
\right]- 4 \sin^2\left[\frac{ 2n\pi}{2(2M+1)}\right]  \right\}=4\cos^2[\alpha(2M+1)].
\end{equation}
\end{lemma}
\begin{proof}
Denote
\begin{equation}\label{gdef}
g(\alpha)=\prod_{m=1}^{2M+1}\left\{4 \sin^2\left[\frac{ (2m-1)\pi}{2(2M+1)}-\alpha
\right]- 4 \sin^2\left[\frac{ 2n\pi}{2(2M+1)}\right]  \right\}.
\end{equation}
Function $g(\alpha)$ is { analytic} in the complex $\alpha$-plane and has the second order zeroes at the
points
\begin{equation}\label{zerg}
\alpha_l=\frac{\pi}{2(2M+1)}+\frac{\pi l}{2M+1}, \quad\quad l=0,\pm 1,\pm 2, \ldots
\end{equation}
It follows from (\ref{zerg}) that the
function
\begin{equation}\label{R}
R(\alpha)=\frac{g(\alpha)}{4\cos^2[\alpha(2M+1)]}
\end{equation}
is { analytic} and has no zeroes in the complex $\alpha$-plane.
Furthermore, this function is rational in the
variable $z=e^{i \alpha}$ and approaches to 1 at $z\to \infty$ and at $z\to 0$.
Therefore, $R(\alpha)=1$.
\end{proof}

Putting $\alpha=0$ in (\ref{l2}) we find
\begin{equation}\label{l20}
\prod_{m=1}^{2M+1}\left\{4 \sin^2\left[\frac{ (2m-1)\pi}{2(2M+1)}
\right]- 4 \sin^2\left[\frac{ 2n\pi}{2(2M+1)}\right]  \right\}=4.
\end{equation}
Since
\begin{eqnarray*}
 \prod_{j=1}^{2M+1}\left\{4 \sin^2\left[\frac{ (2m-1)\pi}{2(2M+1)}
\right]- 4 \sin^2\left[\frac{ 2n\pi}{2(2M+1)}\right]  \right\}=\\
\left[\prod_{m=1}^{M}\left\{4 \sin^2\left[\frac{ (2m-1)\pi}{2(2M+1)}
\right]- 4 \sin^2\left[\frac{ 2n\pi}{2(2M+1)}\right]  \right\}\right]^2\,4\left(1- \sin^2\left[\frac{ 2n\pi}{2(2M+1))}\right]\right),
\end{eqnarray*}
we get
\begin{equation*}
\left[\prod_{m=1}^{M}\left\{4 \sin^2\left[\frac{ (2m-1)\pi}{2(2M+1)}
\right]- 4 \sin^2\left[\frac{ 2n\pi}{2(2M+1)}\right]  \right\}\right]^2=\left[\cos\frac{\pi n}{2M+1}\right]^{-2},
\end{equation*}
or
\begin{equation}\label{pro}
\prod_{m=1}^{M}\left\{4 \sin^2\left[\frac{ (2m-1)\pi}{2(2M+1)}
\right]- 4 \sin^2\left[\frac{ 2n\pi}{2(2M+1)}\right]  \right\}=(-1)^n\left[\cos\frac{\pi n}{2M+1}\right]^{-1}.
\end{equation}
To fix the sign of the right-hand side of (\ref{pro}), we have taken into
account that just the first $n$ factors in the product { on} the left-hand side
are negative at $n=1,\ldots,M$.

Thus,
\begin{eqnarray}\label{2}
\prod_{n=1}^M\prod_{m=1}^M\left[4 \sin^2\frac{ (2m-1)\pi}{2(2M+1)}-
 4 \sin^2\frac{2 n\pi}{2(2M+1)}\right]=
\prod_{n=1}^M\frac{(-1)^{n}}{\cos\frac{\pi n}{2M+1}}.
\end{eqnarray}
To determine the product { on} the right-hand side we use
formula 1.392.1  in  Ref. \cite{Gradshteyn}:
\begin{equation}\label{GR2}
\sin n x=2^{n-1}\prod_{k=0}^{n-1}\sin\left(x+\frac{\pi k}{n}\right).
\end{equation}
For $n=2M+1$, $x=\pi/2$, we get from (\ref{GR2})
\begin{equation*}
\prod_{k=0}^{2M}\cos\left(\frac{\pi k}{2M+1}\right)=2^{-2M}(-1)^M,
\end{equation*}
{ yielding}
\begin{equation}\label{1}
\prod_{k=1}^{M}\cos\left(\frac{\pi k}{2M+1}\right)=2^{-M}.
\end{equation}
Substitution of (\ref{1}) into (\ref{2}) leads finally to (\ref{pr2}).
\renewcommand{\refname}{\centering\textbf{\normalsize References}}

\end{document}